\documentclass[12pt]{iopart}
\usepackage{epsfig}

\newcommand{\tM}{\tilde{\cal M}}
\newcommand{\tu}{\tilde u}
\newcommand{\cR}{{\cal R}}
\newcommand{\cM}{{\cal M}}
\newcommand{\cA}{{\cal A}}
\newcommand{\be}{\begin{equation}}
\newcommand{\ee}{\end{equation}}
\newcommand{\bea}{\begin{eqnarray}}
\newcommand{\eea}{\end{eqnarray}}
\newcommand{\w}[1]{{\bf #1}}
\newcommand{\ie}{{\rm i}}
\newcommand{\nn}{\nonumber \\}
\newcommand{\nicht}[1]{ }
\newcommand{\ex}[1]{{\rm e}^{#1}}

\begin{document}

\title{Rigid Unit Modes in Tetrahedral Crystals}

\author{Franz Wegner}
\address{Institut f\"ur Theoretische Physik,
Universit\"at Heidelberg,\\
Philosophenweg 19, D-69120 Heidelberg}

\begin{abstract}
The 'rigid unit mode' (RUM) model requires unit blocks, in our case tetrahedra
of SiO$_4$ groups, to be rigid within first order of the displacements of the
O-ions.
The wave-vectors of the lattice vibrations, which obey this rigidity, are
determined analytically. Lattices with inversion symmetry yield generically
surfaces of RUMs in reciprocal space, whereas lattices without this symmetry
yield generically lines of RUMs. Only in exceptional cases as in $\beta$-quartz
a surface of RUMs appears, if inversion symmetry is lacking. The occurence of
planes and bending surfaces, straight and bent lines is discussed. Explicit
calculations are performed for five modifications of SiO$_2$ crystals.
\end{abstract}


\section{Introduction}

Displacements of ions in a solid, which alter distances between neighboring ions, produce much stronger forces than those, which vary only angles between adjacing bonds. This has led to the idea of rigid-unit modes
(RUMs), that is distortions, which do not change distances between the ions in a unit to first order in the displacements.
Typical examples are crystals, in which silicon- or aluminium-ions are
surrounded by four oxygen ions. The units of the tetrahedra of these oxygens
are required to be rigid. Each oxygen ion belongs to two tetrahedra.

There have been extensive numerical studies of RUMs of such crystals by Dove,
Giddy, Hammonds and Heine et al. For a review see \cite{Hammonds96}.
More recent presentations of RUMs in framework aluminosilicates can be found in the internet\cite{Dove} and in the review\cite{Dove07}.
These rigid-unit modes do not only signal soft phonon-modes, but they are also at the origin of a large number of displacive phase transitions. The first one considered was the transformation between $\alpha$ and $\beta$ quartz\cite{Grimm75}. Recent applications of RUM modelling include basic ideas for the development of new zeolites\cite{Sartbaeva06}. These zeolites are important catalysts in petrochemical refineries due to their high internal surface areas and molecular sieving properties. Another application deals with
the flexibility of framework structures, which due to RUMs allows cation substitutions with a minimum of energy cost, since the geometric stress associated with the substitution is absorbed by rigid-unit type motion of the polyhedra near the substitution site\cite{Goodwin06}.

Basic to the numerical calculations is the
computer program CRUSH \cite{Giddy93,Hammonds94}. In this program the rigid
tetrahedra are assumed to be individual molecules, and harmonic forces are
added between the two 'split' atoms, which should be one. It calculates the phonon
frequencies $\omega_{j,\w q}$ for given wave-vector $\w q$, which allows to determine a pseudo-intensity
\be
I(\w q) = \sum_j \frac 1{\omega_{j,\w q}^2 + \Omega}
\ee
evaluated for small $\Omega$. For a wave-vector $\w q$ with $n$ RUMs this
quantity approaches $I(\w q)\approx n/\Omega$. In this way the authors
determined RUMs for a large number of crystals.

There are also some analytic calculations along planes and lines of symmetry.
Such a calculation has been performed by Vallade, Berge, and Dolino
\cite{Vallade92} for $\beta$-quartz. The concept of RUMs itself goes back to
Megaw \cite{Megaw71} and Grimm and Dorner \cite{Grimm75}.

The present paper reports analytic calculations of RUMs. The variation
of the lengths of the bonds of the units is calculated as a function of the
displacements of the O-ions. Since the number of coordinates of the
displacements
is the same as the number of lengths of bonds, namely 6 times the number of
tetrahedra, one has to find non-trivial solutions of linear homogeneous
equations, which means the determination of the zeroes of the determinant of
the corresponding coefficient matrix. In reciprocal space these reduce to
$6n_{\rm t}$ equations, where $n_{\rm t}$ is the number of tetrahedra in the
unit cell of the crystal. So one has to
calculate the determinant of a $6n_{\rm t} \times 6n_{\rm t}$ matrix, where the
wave vector $\w q$ enters only through the three complex phase factors
\be
\rho_k = \ex{\ie\w a_k \cdot \w q},
\ee
where $\w a_k$ are the basis vectors of the lattice and $\w q$ is the vector of
the oscillating wave of displacements. We will show that
for a crystal with inversion symmetry the determinant of an appropriately
defined matrix is real for any given $\rho$s. Thus there is only one condition
to be met with three unknowns, which in general defines surfaces of RUMs in
reciprocal space. If
inversion symmetry is lacking, then the determinant is complex, and two
conditions have to be met for the RUMs: both the real and the imaginary part
have to vanish. Correspondingly RUMs are found only at the intersections of the
zeroes of the real and the imaginary part. Thus generically one obtains in
these cases lines of RUMs, but not surfaces. It may happen, however, that both
the real and the imaginary part have a factor in common, so that RUMs extend
over a whole surface. An example is $\beta$-quartz, where a whole plane of
RUMs besides lines of RUMs are found. Although in many cases the RUMs are
located on planes or lines, there are also cases where a surface can bend, as
was already found for HP tridimyte by Dove et al \cite{Dove96}. Here
(eq. \ref{trid}) the analytic expression for the surface is given. Lines of
RUMs may lie in planes. But there are also non-planar lines. Such a line has
been found in $\alpha$-quartz (figure \ref{aqu2b}).

The general theory is given in section \ref{theory}. Some remarks on planes and
bending surfaces, straight and bent lines are given in subsection \ref{bend}.
The basic equations will be derived in subsection \ref{basic}. In particular the
coefficient matrix for the determinant will be given. In subsection
\ref{inversion} the consequences of inversion symmetry will be considered. It
will be shown that the determinant multiplied by appropriate powers of $\rho_k$
is real for lattices with inversion symmetry. This section is concluded with a
few remarks on the actual algebraic calculation in subsection \ref{calc}.

This theory will be applied to five modifications of SiO$_2$ in section
\ref{SiO2}:
first the crystals with inversion symmetry $\beta$-cristobalite and HP tridymite
are considered, then the RUMs for the crystals without inversion symmetry
$\beta$-quartz, $\alpha$-cristobalite, and $\alpha$-quartz are calculated.

\section{General idea\label{theory}}

\subsection{Planes and bending surfaces, straight and bent lines\label{bend}} 

The determinant (or its real and imaginary part) is factorized after
calculation. The factorization procedure of algebraic computer programs is very
useful, since the zeroes of the various factors yield different locations of
RUMs.
Often factors of the form
\be
\rho_1^{l_1}\rho_2^{l_2}\rho_3^{l_3}
\mp \rho_1^{l'_1}\rho_2^{l'_2}\rho_3^{l'_3} \label{rhopr}
\ee
with integer powers $l_i$ and $l'_i$ appear. Then the determinant (or its real
or
imaginary part) vanishes along the plane
\be
[(l_1-l'_1)\w a_1+(l_2-l'_2)\w a_2+(l_3-l'_3)\w a_3] \cdot \w q =
\left\{\begin{array}c 2m\pi \\ (2m+1)\pi \end{array} \right.
\ee
with integer $m$. If we introduce the reciprocal wave-vectors $\w b$ defined by
\be
\w a_i \cdot \w b_j = 2\pi \delta_{ij}
\ee
and represent the wave-vector $\w q$ in this basis,
\be
\w q=\sum_i \xi_i \w b_i = \xi\w b_1 +\eta\w b_2+\zeta\w b_3,
\quad \w a_i \cdot \w q=2\pi\xi_i \label{wK}
\ee
then the plane is given by
\be
(l_1-l'_1)\xi + (l_2-l'_2)\eta + (l_3-l'_3)\zeta = \left\{\begin{array}c m \\
m+\frac 12 \end{array}\right.
\ee
But there are also cases where the factor has a more complex form, which yields
a winding surface or a winding line. Such a surface appears in HP  tridymite,
eq. (\ref{trid}). Lines of
RUMs are planar, if they come from a factor of type (\ref{rhopr}) of the real or
imaginary part. If the factors of both the real and the imaginary part are of
type (\ref{rhopr}) then the line of RUMs is straight. If none of it is of this
type, then the lines are generically non-planar as in $\alpha$-quartz (figure \ref{aqu2b}).

\subsection{Basic equations\label{basic}}

Suppose the oxygens are located at
\bea
\w r_{\w n,i} = \w R_{\w n,i}+\w u_{\w n,i}, \\
\w R_{\w n,i}=\sum_{\alpha} (n_{\alpha} +c_{i\alpha}) \w a_{\alpha}
=\cA (\w n+\w c_i)
\eea
with the equilibrium positions $\w R$ and the displacements $\w u$.
The integers $n_{\alpha}$ number the elementary cells. The atomic coordinates of the $2n_{\rm t}$ oxygens in the elementary cell are denoted by $c_{i,\alpha}$.
The operator $\cA$  maps the unit cube onto the elementary cell. It performs the
similarity transformation
\be
\cA \w x = \sum_{\alpha} \w a_{\alpha} (\w e_{\alpha} \cdot \w x),
\ee
from the orthogonal unit vectors $\w e_{\alpha}$ to the lattice vectors
$\w a_{\alpha}$. Thus the components of $\w x$ and $\w n+\w c$ expand the
vectors in the basis $\{\w a\}$.
The distance between the oxygens at the corners of the tetrahedra should be
fixed in first order in the displacements. Thus for such a pair of atoms at
$\w r_{\w n+\w m,i}$ and $\w r_{\w n+\w m',j}$ with fixed distance one obtains
the condition
\be
(\w R_{\w n+\w m,i}-\w R_{\w n+\w m',j}) \cdot
(\w u_{\w n+\w m,i}-\w u_{\w n+\w m',j}) = \w 0. \label{bestn}
\ee
The distance vector reads
\bea
\w R_{\w n+\w m,i}-\w R_{\w n+\w m',j} = \cA \w d_{\w m-\w m',i,j},
\label{dist}\\
\w d_{\w m-\w m',i,j} = \w c_i + \w m -\w m' -\w c_j.
\eea
Thus eq. (\ref{bestn}) can be rewritten
\bea
\cA \w d_{\w m-\w m',i,j} \cdot (\w u_{\w n+\w m,i}-\w u_{\w n+\w m',j}) \nn
=\w d_{\w m-\w m',i,j} \cdot 
(\cA^{\rm T}\w u_{\w n+\w m,i}-\cA^{\rm T}\w u_{\w n+\w m',j}) \nn
=\w d_{\w m-\w m',i,j} \cdot
(\tilde{\w u}_{\w n+\w m,i}-\tilde{\w u}_{\w n+\w m',j}) = \w 0, \label{bestn2}
\eea
where $\cA^{\rm T}$ is the transposed operator of $\cA$,
\be
\tilde{\w u} := \cA^{\rm T}\w u 
= \sum_{\alpha} \w e_{\alpha} (\w a_{\alpha} \cdot \w u).
\ee
The operator $\cA^{\rm T}$ allows to express
\be
\cA^{\rm T}\w b_{\alpha} = 2\pi\w e_{\alpha}, \quad
\cA^{\rm T}\w q=2\pi\mbox{\boldmath{$\xi$}}.
\ee
Since the vector $\w d$ is independent of $\w n$ the Fourier
transform of equation (\ref{bestn2}) yields
\be
\w d_{\w m-\w m',i,j} \cdot
(\rho_{\w m}\tilde{\w u}_i(\w q) - \rho_{\w m'}\tilde{\w u}_j(\w q)) = 0
\label{bestq}
\ee
with
\bea
\rho_{\w m} &=& \prod_{k=1}^3 \rho_k^{m_k}, \\
\tilde{\w u}_i(\w q) &=& \sum_{\w n} \exp(-\ie\sum_k n_k\w a_k\cdot \w q)
\tilde{\w u}_{\w n,i}.
\eea
The eqs. (\ref{bestq}) constitute a set of linear homogeneous equations for
the
displacements $\w u$. We are looking for those wavevectors $\w q$ which allow
nontrivial solutions of the set of equations (\ref{bestq}). Note that $\cA$ has
disappeared from the eqs. (\ref{bestq}). Thus lattice parameters of the unit
cell do not enter into the calculation.
Nevertheless inversion symmetry and rotational symmetries play still a role.
To facilitate notation the $6n_{\rm t}$ edges of the $n_{\rm t}$
tetrahedra are numbered by $e$. The edge $e$ connects the vertices $i_e$ and
$j_e$. To this edge the triples $\w m_e$, $\w m'_e$ and the distance vector
$\w d_e=\w d_{\w m_e-\w m'_e,i_e,j_e}$ are associated. Then equation
(\ref{bestq}) can be written
\be
\Delta_e := \w d_e \cdot (\rho_{\w m_e}\tilde{\w u}_{i_e}(\w q)
-\rho_{\w m'_e}\tilde{\w u}_{j_e}(\w q)) = 0.
\ee
This set of homogeneous equations in the elongations $\w u$ has non-trivial
solutions, if the determinant $\cM$ of the matrix $M$ with
$6n_{\rm t}\times 6n_{\rm t}$ elements
\be
M_{e,k\alpha}
= \frac{\partial\Delta_e}{\partial \tilde u_{k\alpha}}
= d_{e\alpha} \left(\rho_{\w m_e}\delta_{i_e,k}
- \rho_{\w m'_e}\delta_{j_e,k}\right).
\ee
vanishes. The index $e$ of the edges denotes the rows, the indices
$k$ and $\alpha$ run independently from 1 to $2n_{\rm t}$ and from 1 to 3,
resp., and number the columns of the matrix.

In general the determinant of $M$ yields a complex number as function of the
$\rho_k$.
Thus both the real part and the imaginary part of $\cM$ have to vanish. The
determinant can be expanded
\be
\cM:= \det(M) =\sum_{ijk} \mu_{ijk} \rho_1^i \rho_2^j \rho_3^k
\ee
with real coefficients $\mu_{ijk}$ and integer $i$, $j$, $k$. Thus
$\cM(\rho_1^*,\rho_2^*,\rho_3^*)=\cM^*(\rho_1,\rho_2,\rho_3)$ holds.
Therefore a vanishing $\cM$ for some wave-vector $\w q$ also implies that it
vanishes for $-\w q$.

\subsection{Inversion Symmetry\label{inversion}}

We show that in the case of inversion symmetry $\cM$ multiplied by an
appropriate factor $\rho_{\w m}$ is real. Suppose the center of inversion is
located at
\be
\w R_{\rm I} = \cA\w p = \sum_{\alpha} p^{\alpha} \w a_{\alpha}
\ee
and denote the sublattice obtained by inversion from the sublattice $i$ by
$I(i)$. ($i$ may, but need not be identical to $I(i)$). Then
\be
\w c_i + \w c_{I(i)} = 2\w p + \w l_i
\ee
holds with integer $l^{\alpha}_i=l^{\alpha}_{I(i)}$.
Inversion is also applied to the edges, which generates $I(e)$ from $e$.
Then the corners at the ends of the edges obey
\be
i_{I(e)} = I(i_e), \quad j_{I(e)}=I(j_e).
\ee
Due to this inversion the two distance vectors add up to zero,
\be
\w d_e+\w d_{I(e)} = \w 0
\ee
with
\bea
\w d_e &=& \w c_{i_e} +\w m_e -\w m'_e -\w c_{j_e}, \\
\w d_{I(e)} &=& \w c_{I(i_e)} +\w m_{I(e)} -\w m'_{I(e)} -\w c_{I(j_e)}.
\eea
Therefore one obtains
\be
\w s_e := \w m_e + \w m_{I(e)} + \w l_{i_e}
=\w m'_e + \w m'_{I(e)} + \w l_{j_e},
\ee
which yields
\be
\tilde{\w m}_e + \tilde{\w m}_{I(e)} = \w 0
\ee
with
\be
\tilde{\w m}_e = \w m_e + \frac{\w l_{i_e}-\w s_e}2, \quad
\tilde{\w m}_{I(e)} = \w m_{I(e)} + \frac{\w l_{i_e}-\w s_e}2. \label{mls}
\ee
The matrix $\tilde M$ is introduced by
\bea
M_{e,k\alpha} &=&  \rho_{\w s_e/2} \rho_{-\w l_k/2} \tilde M_{e,k\alpha},
\label{MtM} \\
\tilde M_{e,k\alpha} &=&
d_{e\alpha} \left(\rho_{\tilde{\w m}_e} \delta_{i_e,k}
-\rho_{\tilde{\w m}'_e}\delta_{j_e,k}\right).
\eea
In rows $e$ and $I(e)$ the only non-zero matrix elements of $\tilde M$ are
\bea
\tilde M_{e,i_e\alpha} = d_{e\alpha}\rho_{\tilde{\w m_e}}, &\quad&
\tilde M_{e,j_e\alpha} = -d_{e\alpha}\rho_{\tilde{\w m'_e}}, \\
\tilde M_{I(e),i_e\alpha} = d_{e\alpha}\rho_{\tilde{-\w m_e}}
= \tilde M_{e,i_e\alpha}^*, &\quad&
\tilde M_{I(e),j_e\alpha} = -d_{e\alpha}\rho_{\tilde{-\w m'_e}}
= \tilde M_{e,j_e\alpha}^*
\eea
Therefore exchanging all pairs of rows $e$ and $I(e)$ will transform
the determinant $\tM$ into $(-)^{3n_{\rm t}}$ times the determinant in
which all arguments $\rho_i$ are replaced by $1/\rho_i$ (note that
$\rho_{-\w m}=1/\rho_{\w m}$). Since a tetrahedron
does
not transform into itself under inversion, inversion symmetry is only possible
for even $n_{\rm t}$. Thus
\be
\tM(\rho_1,\rho_2,\rho_3)
= \tM\left(\frac1{\rho_1},\frac1{\rho_2},\frac1{\rho_3}\right). \label{Minv}
\ee
Since for real wave-vectors $\w q$ one has $1/\rho_i=\rho_i^*$, one deduces
\be
\tM=\tM^*.
\ee
Thus $\tM$ is real for crystals with inversion symmetry.
The connection between $\tM$ and $\cM$ is obtained from
eqs. (\ref{mls}, \ref{MtM})
\bea
\tM: = \det(\tilde M) = (\rho_{\Sigma \w m})^{-3/2} \cM,\\
(\rho_{\Sigma \w m})^{-3/2} =
\prod_e (\rho_{\w s_e})^{-1/2} \prod_i (\rho_{\w l_i})^{3/2}.
\eea
The sum $\sum\w m$ has to be extended over all vertices of all tetrahedra. Since
each vertex belongs to two tetrahedra, there are two contributions to each
vertex.

Although relation (\ref{Minv}) holds for crystals with inversion center only,
the transformation from $\cM$ to $\tM$ is also useful for lattices without
inversion symmetry for the following reason: The introduction of the $\w m$ is
to some extend arbitrary. One may add an arbitrary $\w m_{\rm c}$ to $\w m$ at
all corners of a given tetrahedron without changing the distance vectors $\w d$,
$\w m'=\w m+\w m_{\rm c}$. Then in all six rows for the edges of this
tetrahedron an extra factor $\rho_{\w m_{\rm c}}$ appears,
$\cM'=\rho^6_{\w m_{\rm c}} \cM$. Then one obtains
$\sum\w m'=\sum\w m+4\w m_{\rm c}$ and $\tM'=\tM$. Therefore $\tM$ is invariant
against this arbitrary choice in contrast to $\cM$. The over-all sign of $\tM$
depends on the sequence of the vertices. Thus it is arbitrary.

As a consequence crystals with inversion symmetry will show areas for RUMs in
reciprocal space, since only one condition has to be fulfilled. If inversion
symmetry is lacking, then both the real and the imaginary part of $\tM$ has to
vanish. Unless both conditions coincide for whole areas, one obtains only lines
for RUMs in reciprocal space.

\subsection{Remarks on the calculation\label{calc}}

The calculation is performed by means of the algebraic computer program MAPLE.
If the lattice is without inversion
center, then $\tM$ is decomposed into its real $\cR_1$ and its imaginary part
$\cR_2$. If besides the three factors $\rho_k$ there are no other variables,
then the calculation of the determinant is extremely fast. If there is one
extra variable like $x$ in $\beta$-quartz then it takes a few seconds. If there
are three variables, like $x_2$, $y_2$, and $z_2$ in $\alpha$-cristobalite and
in $\alpha$-quartz, then it takes to the order of hours.
However if one assigns rational fractions like $x_2=23976/100000$, then it runs
very fast, whereas it takes quite a while, if one chooses decimal fractions like
$x_2=0.23976$. Apparently the exact calculation facilitates the calculation of
the determinant, although it produces fractions with numerators and denominators
of enormous size. Then also factorization still works, whereas it does not for
decimal fractions due to rounding errors.

The degeneracy of the modes can easily be calculated for RUMs on planes or on
straight lines by determining the rank of the matrix $\tM$, since the
corresponding restriction can easily be evaluated. If surfaces or lines are
bent, then this is more difficult, since the constrained $\rho_k$ have to be
given explicetely. In the following we will not determine the degeneracies.

The crystallographic data used here are those of \cite{navy}. They describe average positions of the ions.
They are almost not representative of the real local structure.
Indeed the average positions are inconsistent with pair distribution functions\cite{Wells04}. This obviously is due to the RUMs. Thus the present analysis gives the RUM spectrum of the average structure.

To the extend the location of the RUMs is given by symmetry, there will be obvious agreement between those determined in refs. \cite{Hammonds96,Dove,Dove07} and ours. For bent lines many agreements will be found. In comparison three things have to be kept in mind:
First of all the cited references give only graphical representations. Therefore comparison is made by appearance only. Secondly, unless the atomic coordinates are determined by symmetry, theirs may differ from the ones used here. Thirdly, apparently Dove et al have included quasi RUMs that is regions in reciprocal space with a low phonon frequency, whereas here only zero frequency modes are taken into account.

\section{Various SiO$_2$ Crystals\label{SiO2}}

In the following the RUMs for five modifications of SiO$_2$ crystals are
determined.

\subsection{$\beta$-Cristobalite}

The coordinates $c'$ for the Si-ions and $c$ for the O-ions are given by
\be \fl
\begin{array}{r|ccc}
i & c'_{i,1} & c'_{i,2} & c'_{i,3} \\ \hline
1 & 1/8 & 1/8 & 1/8 \\
2 & -1/8 & -1/8 & -1/8
\end{array} \qquad
\begin{array}{r|ccc}
i & c_{i,1} & c_{i,2} & c_{i,3} \\ \hline
1 & 0 & 0 & 0 \\
2 & 1/2 & 0 & 0 \\
3 & 0 & 1/2 & 0 \\
4 & 0 & 0 & 1/2
\end{array} \qquad
\begin{array}{r@{\,=\,}l}
\w a_1 & \frac a2 \w e_2 + \frac a2 \w e_3 \\
\w a_2 & \frac a2 \w e_1 + \frac a2 \w e_3 \\
\w a_3 & \frac a2 \w e_1 + \frac a2 \w e_2
\end{array}
\ee
The corners of the two tetrahedra are obtained by applying $\cA$ to\\
1st tetrahedron: $\w c_1$, $\w c_2$, $\w c_3$, $\w c_4$;\\
2nd tetrahedron: $\w c_1$, $\w c_2-\w e_1$, $\w c_3-\w e_2$, $\w c_4-\w e_3$.\\
The determinant evaluates to
\be
\tM =
\frac{(1-\rho_1)(1-\rho_2)(1-\rho_3)(\rho_1-\rho_2)(\rho_1-\rho_3)(\rho_2-\rho_3)}
{2^{12}\rho_1^{3/2}\rho_2^{3/2}\rho_3^{3/2}}.
\ee
With eq. (\ref{wK}) one obtains
\bea
\frac{1-\rho_k}{\rho_k^{1/2}} &=& -2\ie\sin(\w a_k\w q/2) = -2\ie\sin(\pi\xi_k),
\\
\frac{\rho_k-\rho_l}{(\rho_k\rho_l)^{1/2}} &=& 2\ie\sin((\w a_k-\w a_l)\w q/2)
= 2\ie\sin(\pi(\xi_k-\xi_l)).
\eea
Thus all RUMs are located in the planes in reciprocal space
\be
(0,\eta,\zeta), \quad (\xi,0,\zeta), \quad (\xi,\eta,0), \quad
(\xi,\xi,\zeta), \quad (\xi,\eta,\xi), \quad (\xi,\eta,\eta).
\ee
The crystal has a center of
inversion. One easily checks eq. (\ref{Minv}) for this crystal.

\paragraph{Comparison} The planes of RUMs found agree completely with those
given in ref. \cite{Hammonds96}. 

\subsection{HP tridymite}

The atomic coordinates of the HP tridymite phase of SiO$_2$ are given by
\be \fl
\begin{array}{r|ccc}
i & c'_{i,1} & c'_{i,2} & c'_{i,3} \\ \hline
1 & 1/3 & 2/3 & z \\
2 & 2/3 & 1/3 & -z \\
3 & 2/3 & 1/3 & 1/2+z \\
4 & 1/3 & 2/3 & 1/2-z
\end{array} \qquad
\begin{array}{r|ccc}
i & c_{i,1} & c_{i,2} & c_{i,3} \\ \hline
1 & 1/3 & 2/3 & 1/4 \\
2 & 2/3 & 1/3 & 3/4 \\
3 & 1/2 &  0 & 0 \\
4 & 0 & 1/2 & 0 \\
5 & 1/2 & 1/2 & 0 \\
6 & 1/2 & 0 & 1/2 \\
7 & 0 & 1/2 & 1/2 \\
8 & 1/2 & 1/2 & 1/2 
\end{array} \qquad
\begin{array}{r@{\,=\,}l}
\w a_1 & \frac 12 a\w e_1 - \frac{\sqrt 3}2 a\w e_2 \\
\w a_2 & \frac 12 a\w e_1 + \frac{\sqrt 3}2 a\w e_2 \\
\w a_3 & c\w e_3 \label{tridc}
\end{array}
\ee
The eqs. (\ref{bestq}) have to be fulfilled for the edges of the four
tetrahedra with corners obtained by applying $\cA$ to\\
1st tetrahedron: $\w c_1$, $\w c_3+\w e_2$, $\w c_4$, $\w c_5$;\\
2nd tetrahedron: $\w c_2-\w e_3$, $\w c_3$, $\w c_4+\w e_1$, $\w c_5$;\\
3rd tetrahedron: $\w c_2$, $\w c_6$, $\w c_7+\w e_1$, $\w c_8$;\\
4th tetrahedron: $\w c_1$, $\w c_6+\w e_2$, $\w c_7$, $\w c_8$.\\
The determinant $\tM$ yields
\bea
\tM &=& f^2_2(\rho_1,\rho_2)\frac{(1-\rho_3)}{2^{32}3^2\rho_3^{1/2}}
\cR,\label{tmHP}\\
\cR &=& 9\left(\rho_3+\frac 1{\rho_3}\right)
-4f_1(\rho_1,\rho_2)+14,\\
f_1(\rho_1,\rho_2) &=& \frac{(1+\rho_1)(1+\rho_2)(1+\rho_1\rho_2)}{\rho_1\rho_2}
\nn
&=& 8\cos(\w a_1\w q/2)\cos(\w a_2\w q/2)\cos((\w a_1+\w a_2)\w q/2) \nn
&&=2\big(1+\cos(\w a_1\w q)+\cos(\w a_2\w q)+\cos((\w a_1+\w a_2)\w q)\big),\\
f_2(\rho_1,\rho_2) &=& \frac{(1-\rho_1)(1-\rho_2)(1-\rho_1\rho_2)}{\rho_1\rho_2}
\nn
&=& 8\ie\sin(\w a_1\w q/2)\sin(\w a_2\w q/2)\sin((\w a_1+\w a_2)\w q/2)\nn
&&=2\ie\big(\sin(\w a_1\w q)+\sin(\w a_2\w q)-\sin((\w a_1+\w a_2)\w q)\big).
\eea
These functions $f_1$ and $f_2$ as well as the later introduced functions $f_3$,
eq. (\ref{f3}), and $f_4$, eq. (\ref{f4}) are invariant under rotations by
$2\pi/3$ around the $z$-axis, which transforms
\bea
\w a_1 \rightarrow \w a_2 \rightarrow -\w a_1-\w a_2 \rightarrow \w a_1, \nn
\rho_1 \rightarrow \rho_2 \rightarrow 1/(\rho_1\rho_2) \rightarrow \rho_1.
\eea

Due to the factor $1-\rho_3$ in eq. (\ref{tmHP}) RUMs are located in the plane
\be
(\xi,\eta,0).
\ee
The factors of $f_2$ yield RUMs in the planes
\be
(0,\eta,\zeta), \quad (\xi,0,\zeta), \quad (\xi,-\xi,\zeta).
\ee
In contrast the zeroes of $\cR$ describe a winding surface in reciprocal space,
which may be written
\bea
\cos(2\pi\zeta) &=& \frac{16}9 \cos(\pi\xi)\cos(\pi\eta)\cos(\pi(\xi+\eta))
-\frac 79 \nn
&=& \frac 49 \left(\cos(2\pi\xi)+\cos(2\pi\eta)
+\cos(2\pi(\xi+\eta))\right) - \frac 13. \label{trid}
\eea
Note that the maximum of the sum of the r.h.s is $+1$, which is obtained at
[0,0,0]. The minimum of the r.h.s. is $-1$, which is reached at
$[\pm 1/3,\pm 1/3,1/2]$. For $\xi=1/2$ or $\eta=1/2$ or $\xi+\eta=1/2$ one
obtains $\cos(2\pi\zeta)=-7/9$, which yields $\zeta=l=\pm 0.39183$. 

\paragraph{Comparison} These results are in full agreement with those obtained
by Dove et al. \cite{Dove96} by means of their numerical CRUSH-program
\cite{Giddy93,Hammonds94} and reported in ref. \cite{Hammonds96}.

\subsubsection*{Other derivation}

One may determine this bending RUM also in the following way:
One starts with the equations for the edges between $\w c_3$, $\w c_4$, and $\w
c_5$.
Since they all lie in the $xy$-plane, the third component does not enter and
one obtains the equations
\bea
\tu_{4,1} = \tu_{5,1}, &&\qquad \rho_1 \tu_{4,1}=\tu_{5,1}, \nn
\tu_{3,2} = \tu_{5,2}, &&\qquad \rho_2 \tu_{3,2}=\tu_{5,2}, \nn
\rho_2 (\tu_{3,1}+\tu_{3,2}) = \tu_{4,1}+\tu_{4,2}, &&\qquad
\tu_{3,1}+\tu_{3,2} = \rho_1 (\tu_{4,1}+\tu_{4,2}).
\eea
Evidently, if
\be
\rho_1\not=1, \quad \rho_2\not=1, \quad \rho_1\rho_2\not=1, \label{not}
\ee
then all these components vanish
\be
\tu_{3,1}=\tu_{3,2}=\tu_{4,1}=\tu_{4,2}=\tu_{5,1}=\tu_{5,2}=0.
\ee
Similarly one shows by considering the third and fourth tetrahedron that under
the same condition (\ref{not}) one obtains
\be
\tu_{6,1}=\tu_{6,2}=\tu_{7,1}=\tu_{7,2}=\tu_{8,1}=\tu_{8,2}=0.
\ee
This implies that the tetrahedra are rotated around axes parallel to the
xy-plane. There are twelve equations left for the twelve other components $\tu$.
One can use six of them to eliminate $\tu_{3,3}$ to $\tu_{8,3}$.
Finally one calculates the determinant of the coefficient matrix
\be
\left(\begin{array}{cccccc}
-\frac{\rho_3}{6\rho_2} & -\frac{\rho_3}{3\rho_2} & \frac{\rho_3}{4\rho_2} &
\frac 16 & \frac 13 & -\frac 14 \\
\frac{\rho_1\rho_3}3 & \frac{\rho_1\rho_3}6 & \frac{\rho_1\rho_3}4 &
-\frac 13 & -\frac 16 & -\frac 14 \\
-\frac{\rho_3}6 & \frac{\rho_3}6 & \frac{\rho_3}4 &
\frac 16 & -\frac 16 & -\frac 14 \\
-\frac 1{6\rho_2} & -\frac 1{3\rho_2} & -\frac 1{4\rho_2} &
\frac 16 & \frac 13 & \frac 14 \\
\frac{\rho_1}3 & \frac{\rho_1}6 & -\frac{\rho_1}4 &
-\frac 13 & -\frac 16 & \frac 14 \\
-\frac 16 & \frac 16 & -\frac 14 & \frac 16 & -\frac 16 & \frac 14
\end{array}\right) 
\ee 
of the last six
equations for the six components of $\tilde{\w u}_1$ and $\tilde{\w u}_2$, which
has to vanish. This determinant factorizes in a factor $\rho_3-1$ and into
\bea
&& 2\left(\frac 1{\rho_1} + \rho_1 + \frac 1{\rho_2} + \rho_2 + \frac
1{\rho_1\rho_2}
+ \rho_1\rho_2\right) - \frac 92 \left(\frac 1{\rho_3}+\rho_3\right) -3 \nn
&=& 4\big(\cos(\w a_1\w q)+\cos(\w a_2\w q)+\cos((\w a_1+\w a_2)\w q)\big)
-9 \cos(\w a_3\w q) - 3 =0,
\eea
which again yields eq. (\ref{trid}).

\subsection{$\beta$-Quartz}

\begin{figure}[ht]
\centerline{\epsfig{file=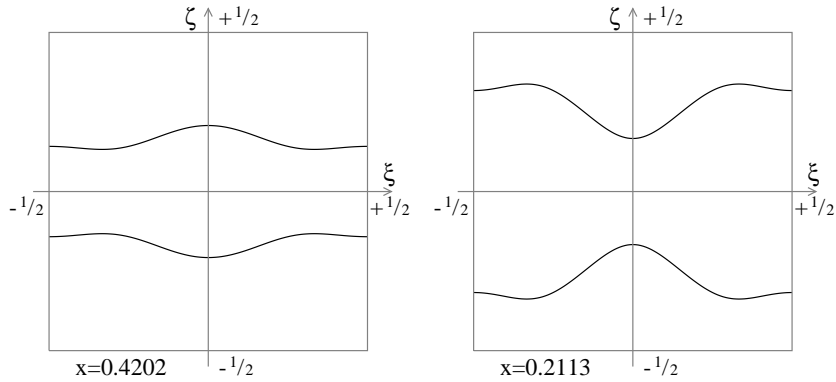}}
\caption{$\zeta(\xi)$ of the RUMs  of $\beta$-quartz given by eqs. (\ref{bquxz},
\ref{bquxz1}) for $x=0.4202$ and $x=0.2113$.\label{bqu}}
\end{figure}

The coordinates for the Si and O-ions are given by
\be \fl
\begin{array}{r|ccc}
i & c'_{i,1} & c'_{i,2} & c'_{i,3} \\ \hline
1 & 1/2 & 0 & 0 \\
2 & 0 & 1/2 & 2/3 \\
3 & 1/2 & 1/2 & 1/3
\end{array} \qquad
\begin{array}{r|ccc}
i & c_{i,1} & c_{i,2} & c_{i,3} \\ \hline
1 & x & 2x & 1/2 \\
2 & -2x & -x & 1/6 \\
3 & x & -x & 5/6 \\
4 & -x & -2x & 1/2 \\
5 & 2x & x & 1/6 \\
6 & -x & x & 5/6
\end{array} \qquad
\begin{array}{r@{\,=\,}l}
\w a_1 & \frac 12 a\w e_1 - \frac{\sqrt 3}2 a\w e_2 \\
\w a_2 & \frac 12 a\w e_1 + \frac{\sqrt 3}2 a\w e_2 \\
\w a_3 & c\w e_3
\end{array}
\ee
The corners of the three tetrahedra are obtained by applying $\cA$ to\\
1st tetrahedron: $\w c_2+\w e_1$, $\w c_3-\w e_3$, $\w c_5$,
$\w c_6+\w e_1-\w e_3$;\\
2nd tetrahedron: $\w c_1$, $\w c_3+\w e_2$, $\w c_4+\w e_2$, $\w c_6$;\\
3rd tetrahedron: $\w c_1$, $\w c_2+\w e_1+\w e_2$, $\w c_4+\w e_1+\w e_2$,
$\w c_5$.\\
The determinant reads
\bea
\tM &=& \left(\frac{4x(3x-1)}9\right)^3 \frac{1-\rho_3}{\rho_3^{1/2}}
\left(\cR_1 + \cR_2\right), \\
\cR_1 &=& \frac{1+\rho_3}{\rho_3^{1/2}}f_2(\rho_1,\rho_2) \nn
&\times& \left(-k_3^3\left(\frac{1+\rho_3^2}{\rho_3}\right)+k_1+k_2
f_1(\rho_1,\rho_2)\right), \\
\cR_2 &=& 2x(2x-1)(3x-1)(4x-1)(6x-1)\frac{1-\rho_3}{\rho_3^{1/2}}
f_3(\rho_1,\rho_2), \\
f_3(\rho_1,\rho_2) &=&
\frac{(\rho_1-\rho_2)(1-\rho_1\rho_2^2)(1-\rho_1^2\rho_2)}{\rho_1^2\rho_2^2},\\
k_1 &=& 2x(4x-1)(432x^4-540x^3+252x^2-51x+4), \label{f3} \\
k_2 &=& -x(4x-1)(3x-1)^2, \\
k_3 &=& 12x^2-6x+1.
\eea
Due to the factor $(1-\rho_3)$ RUMs are obtained in the plane
\be
(\xi,\eta,0).
\ee
The factor $(1+\rho_3)$ in $\cR_1$ and the zeroes of $f_3$ in
$\cR_2$ yield RUMs along the straight lines
\be
[\xi,\xi,1/2], \quad [\xi,-2\xi,1/2], \quad [-2\xi,\xi,1/2].
\ee
The zeroes of $f_2$ in $\cR_1$ and of $f_3$ in
$\cR_2$ yield RUMs along the straight lines
\be
[0,0,\zeta], \quad [0,1/2,\zeta], \quad [1/2,0,\zeta], \quad [1/2,1/2,\zeta].
\ee
Finally the zeroes of the large parenthesis in the expression for
$\cR_1$ and the zeroes of $f_3$ in $\cR_2$ yield 
RUMs along the curves
\be
[\xi,\xi,\zeta(\xi)], \quad [\xi,-2\xi,\zeta(\xi)], \quad
[-2\xi,\xi,\zeta(\xi)], \label{bquxz}
\ee
where $\zeta(\xi)$ is given by
\be
\cos(2\pi\zeta(\xi)) = \frac{k_1+8k_2\cos^2(\pi\xi)\cos(2\pi\xi)} {2k_3^3}.
\label{bquxz1}
\ee
With $x=0.4202$ of ref. \cite{navy} one obtains
\be
k_1= 0.26799, \quad k_2= -0.019427, \quad k_3= 0.59762
\ee
and
\be
\cos(2\pi\zeta(\xi)) = 0.6278-0.3641 \cos^2(\pi\xi)\cos(2\pi\xi),
\ee
However, this value of $x$ yields tetrahedra, which are far from being
equilateral. The value for equilateral tetrahedra is
\be
x=\frac 12-\frac 1{\sqrt{12}} = 0.2113, \label{bqux}
\ee
which yields
\bea
k_1= -0.015801, \quad k_2= 0.004380, \quad k_3= 0.26795, \\
\cos(2\pi\zeta(\xi)) = -0.41068+0.91068 \cos^2(\pi\xi)\cos(2\pi\xi).
\eea
The function $\zeta(\xi)$ is plotted for both values $x$ in fig. \ref{bqu}.

\paragraph{Comparison} We find agreement of the RUMs for $\beta$-quartz in
reciprocal space with those given in refs. \cite{Dove96,Hammonds96,Dove,Dove07}. This includes the winding lines of fig. \ref{bqu} seen in ref. \cite{Dove,Dove07} with the ideal value $x$ of eq. (\ref{bqux}).

\subsection{$\alpha$-Cristobalite}

The coordinates $c'$ for the Si-ions and $c$ for the O-ions are given by
\be \fl
\begin{array}{r|ccc}
i & c'_{i,1} & c'_{i,2} & c'_{i,3} \\ \hline
1 & x_1 & x_1 & 0 \\
2 & -x_1 & -x_1 & 1/2 \\
3 & 1/2-x_1 & 1/2+x_1 & 1/4 \\
4 & 1/2+x_1 & 1/2-x_1 & 3/4
\end{array} \qquad
\begin{array}{r|ccc}
i & c_{i,1} & c_{i,2} & c_{i,3} \\ \hline
1 & x_2 & y_2 & z_2 \\
2 & -x_2 & -y_2 & 1/2+z_2 \\
3 & 1/2-y_2 & 1/2+x_2 & 1/4+z_2 \\
4 & 1/2+y_2 & 1/2-x_2 & 3/4+z_2 \\
5 & y_2 & x_2 & -z_2 \\
6 & -y_2 & -x_2 & 1/2-z_2 \\
7 & 1/2-x_2 & 1/2+y_2 & 1/4-z_2 \\
8 & 1/2+x_2 & 1/2-y_2 & 3/4-z_2 
\end{array} \qquad
\begin{array}{r@{\,=\,}l}
\w a_1 & a\w e_1 \\
\w a_2 & a\w e_2 \\
\w a_3 & c\w e_3
\end{array}
\ee
The corners of the four tetrahedra are obtained by applying $\cA$ to\\
1st tetrahedron: $\w c_1$, $\w c_4-\w e_3$, $\w c_5$, $\w c_7$;\\
2nd tetrahedron: $\w c_2$, $\w c_3-\w e_1-\w e_2$, $\w c_6$,
$\w c_8-\w e_1-\w e_2$;\\ 
3rd tetrahedron: $\w c_1+\w e_2$, $\w c_3$, $\w c_6+\w e_2$, $\w c_7$;\\
4th tetrahedron: $\w c_2+\w e_1$, $\w c_4$, $\w c_5+\w e_1+\w e_3$, $\w c_8$.\\
The determinant reads
\bea
\tM &=& k_0^4(\cR_1+\cR_2), \\
\cR_1 &=& \sum_{l=0}^2 g_l \cos^l(2\pi\xi_3),\\
\cR_2 &=& k' \frac{(1-\rho_1^2)(1-\rho_2^2)(1-\rho_3^2)(\rho_1-\rho_2)
 (1-\rho_1\rho_2)}{\rho_1^2\rho_2^2\rho_3}, \\
g_l &=& \sum_{ij} k_{ijl} \cos^i(2\pi\xi_1) \cos^j(2\pi\xi_2), \\
k_0 &=& \frac 18 (2x_2-1)(8z_2y_2+x_2-y_2)
\eea
with
\be
k_{ijl} = k_{jil}.
\ee
The sum over $i$ and $j$ runs at most up to $i=3$, $j=3$, $i+j=4$.
The coefficients $k_{ijl}$ are polynomials in $x_2$, $y_2$, $z_2$. They are of
order 4 in $z_2$ and run up to order 8 in $x_2$ and $y_2$. Thus in general they
are lengthy expressions.

For
\be
x_2=0.23976, \quad y_2=0.10324, \quad z_2=0.17844 \label{Crxyz}
\ee
from ref. \cite{navy} one obtains
\be
k'=-0.00052747, \quad k_0=-0.018470,
\ee
and the non-zero cofficients $k_{ijl}$
\be
\begin{array}{rr|r}
i & j & 100 k_{ij0} \\ \hline
0 & 0 & -8.0953 \\
0 & 1 & 9.6093 \\
0 & 2 & 9.4540 \\
0 & 3 & -10.9503 \\
1 & 1 & -8.4081 \\
1 & 2 & -7.3564 \\
1 & 3 & 7.9854
\end{array}
\qquad
\begin{array}{rr|r}
i & j & 100 k_{ij1} \\ \hline
0 & 0 & -0.7104 \\
0 & 1 & -3.7422 \\
0 & 2 & 2.5622 \\
1 & 1 & -19.5336 \\
1 & 2 & 18.3067 \\
2 & 2 & -15.9707
\end{array}
\qquad
\begin{array}{rr|r}
i & j & 100 k_{ij2} \\ \hline
0 & 0 & 8.8057 \\
0 & 1 & -5.8672 \\
1 & 1 & 3.9092
\end{array}
\ee

\begin{figure}[ht]
\centerline{\epsfig{file=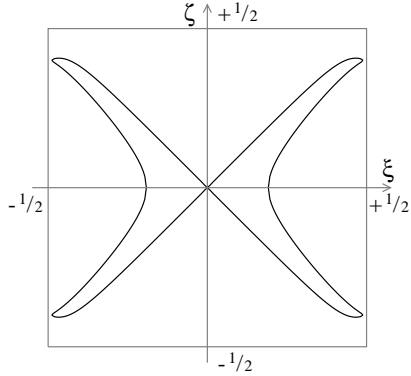}}
\caption{$\zeta(\xi)$ of the RUMs of $\alpha$-cristobalite described by
eqs.(\ref{alCr1}, \ref{alCr1b}, \ref{alCr1p}).\label{cr1p}}
\end{figure}

The zeroes $\rho_1=\pm 1$, $\rho_2=\pm 1$, $\rho_2=\rho_1$, $\rho_2=1/\rho_1$ of
$\cR_2$ yield RUMs for the zeroes of $\cR_1$ from
\be
\cos(2\pi\zeta(\xi))
= \frac{-g_1\pm\sqrt{d}}{2g_2}, \quad d=g_1^2-4g_0g_2. \label{alCr1}
\ee
For $\rho_1=1$ and $\rho_2=1$ one obtains RUMs along lines
\be
[0,\xi,\zeta(\xi)], \quad [\xi,0,\zeta(\xi)], \label{alCr1b}
\ee
resp. with
\bea
10^2 g_2 &=& 2.9385-1.9579 \cos(2\pi\xi), \nn
10^2 g_1 &=& -1.8904 -4.9691 \cos(2\pi\xi) + 4.8982 \cos^2(2\pi\xi), \nn
10^4 d &=& (1-\cos(2\pi\xi))^2 (3.3647 +4.1431 \cos(2\pi\xi) +0.7718
\cos^2(2\pi\xi)). \label{alCr1p}
\eea
They are shown in fig. \ref{cr1p}.

For $\rho_1=-1$ and $\rho_2=-1$ one obtains RUMs along lines
\be
[1/2,\xi,\zeta(\xi)], \quad [\xi,1/2,\zeta(\xi)]
\ee
resp. with
\bea
10^2 g_2 &=& 14.6729 -9.7764 \cos(2\pi\xi), \nn
10^2 g_1 &=& 5.5940 + 34.0981 \cos(2\pi\xi) - 31.7152 \cos^2(2\pi\xi), \nn
10^4 d &=& - 127.156 + 330.023 \cos(2\pi\xi) -74.142 \cos^2(2\pi\xi) \nn
&&  - 394.118 \cos^3(2\pi\xi) + 265.367 \cos^4(2\pi\xi).
\eea
The calculation with (\ref{Crxyz}) did not give real solutions $\zeta(\xi)$.

\begin{figure}[ht]
\centerline{\epsfig{file=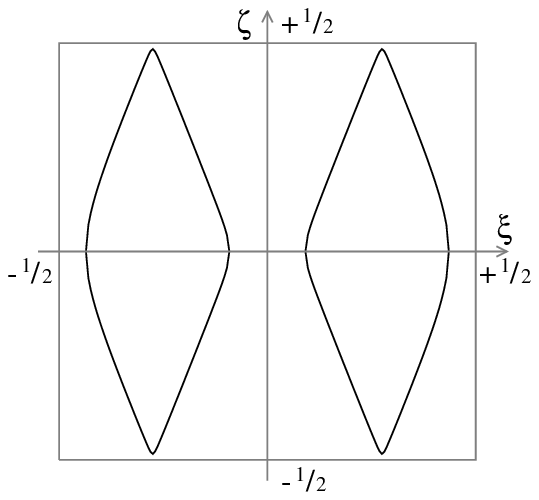}}
\caption{$\zeta(\xi)$ of the RUMs of $\alpha$-cristobalite given by eqs.
(\ref{alCr3b}, \ref{alCr3}, \ref{alCr3a}).\label{cr1z}}
\end{figure}

For $\rho_2=\rho_1$ and $\rho_2=1/\rho_1$ one obtains RUMs along lines
\be
[\xi,\xi,\zeta(\xi)], \quad [\xi,-\xi,\zeta(\xi)] \label{alCr3b}
\ee
with
\bea
10^2 g_2 &=& (2.9674 - 1.9772 \cos(2\pi\xi))^2, \nn
10^2 g_1 &=& - 0.7104 - 7.4844 \cos(2\pi\xi)  - 14.4092 \cos^2(2\pi\xi) \nn
&& + 36.6135 \cos^3(2\pi\xi)  - 15.9707 \cos^4(2\pi\xi), \nn
d &=& (2g_2+g_1)^2. \label{alCr3}
\eea
The square root of $d$ is rational in this case. The first solution
(\ref{alCr1}) yields the
straight line of RUMs
\be
[\xi,\pm\xi,0], \label{alCr2}
\ee
The second one yields the relation
\be
\cos(2\pi\zeta(\xi)) = -1-\frac{g_1}{g_2}. \label{alCr3a}
\ee
The locations of these RUMs in reciprocal space are shown in fig. \ref{cr1z}.

\begin{figure}[ht]
\centerline{\epsfig{file=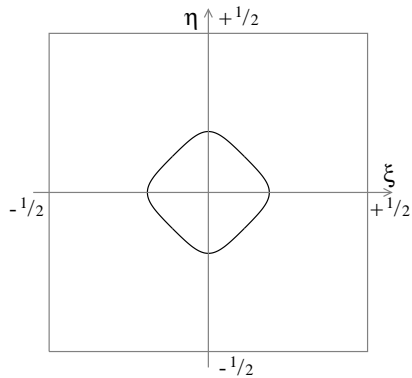}}
\caption{$\eta$ versus $\xi$ of the RUMs of $\alpha$-cristobalite given by eqs.
(\ref{alCr4}, \ref{alCr4a}).\label{cr2}}
\end{figure}

The RUMs given by the zero $\rho_3=1$ are the RUMs eq. (\ref{alCr2}) just
mentioned plus RUMs
along the lines $[\xi,\eta,0]$ with $\xi$ and $\eta$ related by
\be
\cos(2\pi\xi)\cos(2\pi\eta) - A (\cos(2\pi\xi) + \cos(2\pi\eta))+B=0,
\label{alCr4}
\ee
where for the above given $x_2$, $y_2$, $z_2$ one obtains
\be
A=1.3713, \quad B=1.5048. \label{alCr4a}
\ee
The corresponding line is plotted in fig. \ref{cr2}.

Finally the RUMs given by the zero $\rho_3=-1$ of $\cR_2$ are given by lines
$[\xi,\eta,1/2]$ with $\xi$ and $\eta$ related by
\be
\sum_{i,j=0}^3 k_{i,j}\cos^i(2\pi\xi) \cos^j(2\pi\eta)=0,
\ee
which with (\ref{Crxyz}) yields
\bea
&& 15.9707 \cos^2(2\pi\xi) \cos^2(2\pi\eta) \nn
&& +7.9854 \cos(2\pi\xi) \cos(2\pi\eta) (\cos^2(2\pi\xi)+\cos^2(2\pi\eta)) \nn
&& -25.6631 \cos(2\pi\xi) \cos(2\pi\eta) (\cos(2\pi\xi)+\cos(2\pi\eta)) \nn
&& -10.9503 (\cos^3(2\pi\xi)+\cos^3(2\pi\eta)) \nn
&& +15.0347 \cos(2\pi\xi) \cos(2\pi\eta)
+6.8918(\cos^2(2\pi\xi)+\cos^2(2\pi\eta)) \nn
&& +7.4844 (\cos(2\pi\xi)+\cos(2\pi\eta)) +1.4208=0.
\eea
This can be rewritten
\be
c_{\rm d}^2 = \frac{c_{\rm s}^4 - 4.5851c_{\rm s}^3 + 3.6089 c_{\rm s}^2 +
3.7491 c_{\rm s} + 0.7117}
{c_{\rm s}^2 + 0.9001 c_{\rm s} + 0.1567} \label{alCr5}
\ee
with
\be
c_{\rm s} = \cos(2\pi\xi)+\cos(2\pi\eta), \quad
c_{\rm d} = \cos(2\pi\xi)-\cos(2\pi\eta).
\ee
The numerical calculation showed that either $c^2_{\rm d}$ from eq.
(\ref{alCr5}) is negative or one of the cosines is larger than one. Therefore
there is no real solution of this type.

\paragraph{Comparison} The figure for $\alpha$-cristobalite in ref. \cite{Dove,Dove07}
shows the RUMs of fig. \ref{cr1p} and \ref{cr2} and the line eq. \ref{alCr2}
(green) for $\xi,\eta,\zeta\ge 0$. The line fig. \ref{cr1z} cannot be seen.
However, a region of quasi-RUMs (modes of small frequency) is shown, which
probably hides the line of fig. \ref{cr1z}.

\subsection{$\alpha$-Quartz}

The coordinates of the Si- and O-ions in $\alpha$-quartz are
\be \fl
\begin{array}{r|ccc}
i & c'_{i,1} & c'_{i,2} & c'_{i,3} \\ \hline
1 & x_1 & 0 & 2/3 \\
2 & 0 & x_1 & 1/3 \\
3 & -x_1 & -x_1 & 0
\end{array} \qquad
\begin{array}{r|ccc}
i & c_{i,1} & c_{i,2} & c_{i,3} \\ \hline
1 & x_2 & y_2 & z_2 \\
2 & -y_2 & x_2-y_2 & 2/3+z_2 \\
3 & y_2-x_2 & -x_2 & 1/3+z_2 \\
4 & y_2 & x_2 & -z_2 \\
5 & -x_2 & y_2-x_2 & 2/3-z_2 \\
6 & x_2-y_2 & -y_2 & 1/3-z_2
\end{array} \qquad
\begin{array}{r@{\,=\,}l}
\w a_1 & \frac 12 a\w e_1 - \frac{\sqrt 3}2 a\w e_2 \\
\w a_2 & \frac 12 a\w e_1 + \frac{\sqrt 3}2 a\w e_2 \\
\w a_3 & c\w e_3
\end{array}
\ee
The corners of the three tetrahedra are obtained by applying $\cA$ to\\
1st tetrahedron: $\w c_1$, $\w c_2+\w e_1-\w e_3$, $\w c_5+\w e_1+\w e_3$,
$\w c_6+\w e_3$;\\
2nd tetrahedron: $\w c_2-\w e_3$, $\w c_3+\w e_2-\w e_3$, $\w c_4+\w e_3$,
$\w c_6+\w e_2+\w e_3$;\\
3rd tetrahedron: $\w c_1-\w e_3$, $\w c_3+\w e_1+\w e_2-\w e_3$,
$\w c_4+\w e_3$, $\w c_5+\w e_1+\w e_2$.\\
The determinant evaluates to
\bea
\tM &=& k_1^3(\cR_1+\cR_2), \\
k_1 &=& \frac{2(3x_2-2)(-2x_2+3z_2x_2-6y_2z_2+5y_2)}9, \\
\cR_1 &=& f_2(\rho_1,\rho_2)\frac{1-\rho^2_3}{\rho_3} 
\left(k_2 f_1(\rho_1,\rho_2)+ k_3 \frac{1+\rho_3^2}{\rho_3} + k_4\right)\\ 
\cR_2 &=& \frac{1-\rho_3^4}{\rho_3^2}(k_5 f_1(\rho_1,\rho_2)
+k_6) \nn
&+& \frac{1-\rho_3^2}{\rho_3}(k_7 f_1(\rho_1,\rho_2)
+k_8 f_1(\rho_1^2,\rho_2^2) + k_9 f_4(\rho_1,\rho_2) +k_{10}) \nn
&+& (k_{11}\frac{1+\rho_3^2}{\rho_3} +k_{12}) f_3(\rho_1,\rho_2), \\
f_4(\rho_1,\rho_2)
&=& \frac{(\rho_1+\rho_2)(1+\rho_1\rho_2^2)(1+\rho_1^2\rho_2)}
{\rho_1^2\rho_2^2}. \label{f4}
\eea
with lengthy polynomial expressions of $x_2$, $y_2$, $z_2$ for the constants
$k_2$ to $k_{12}$.
With
\be
x_2= 0.4141, \quad y_2= 0.2681, \quad z_2= 0.7854
\ee
from ref. \cite{navy} one obtains
\be
\begin{array}{lll}
k_1 = -0.037819, & k_2 = 0.018997, & k_3 = -0.030172, \\
k_4 = 0.000088, & k_5 = 0.006248, & k_6 = -0.060332, \\
k_7 = 0.021768, & k_8 = -0.009082, & k_9 = -0.000285, \\
k_{10} = -0.078514, & k_{11} = -0.006089, & k_{12} = 0.001074.
\end{array}
\ee

\begin{figure}[ht]
\centerline{\epsfig{file=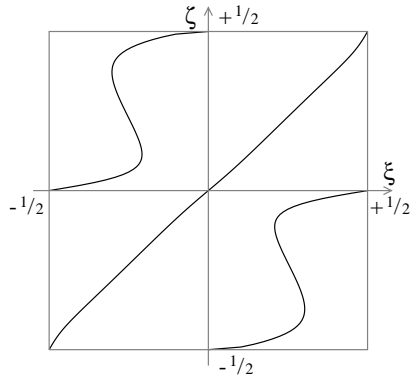}}
\caption{$\zeta(\xi)$ of the RUMs  of $\alpha$-quartz described by eqs.
(\ref{aqu11}, \ref{aqu12}).\label{aqu1}}
\end{figure}

From the zeroes of $\cR_1$ combined with the zeroes of $\cR_2$ one obtains three
classes of solutions:

(i) $f_2(\rho_1,\rho_2)=0$ yields a class of RUMs
\be
[\xi,0,\zeta(\xi)], \quad [0,-\xi,\zeta(\xi)], \quad [-\xi,\xi,\zeta(\xi)]
\label{aqu11}
\ee
where $\zeta(\xi)$ is determined by
\be
\cR_2(\rho_1=\ex{2\pi\ie\xi},\rho_2=1,\rho_3=\ex{2\pi\ie\zeta(\xi)})=0.
\label{aqu12}
\ee
The solutions are shown in fig. \ref{aqu1}. One line of RUMs is nearly straight,
the other one is snaky.

(ii) $1-\rho_3^2=0$ and $f_3(\rho_1,\rho_2)=0$ yield straight lines of RUMs
\bea
&& [\xi,\xi,0], \quad [\xi,-2\xi,0], \quad [-2\xi,\xi,0], \label{aqst0}\\
&& [\xi,\xi,1/2], \quad [\xi,-2\xi,1/2], \quad [-2\xi,\xi,1/2].
\eea

\begin{figure}[ht]
\centerline{\epsfig{file=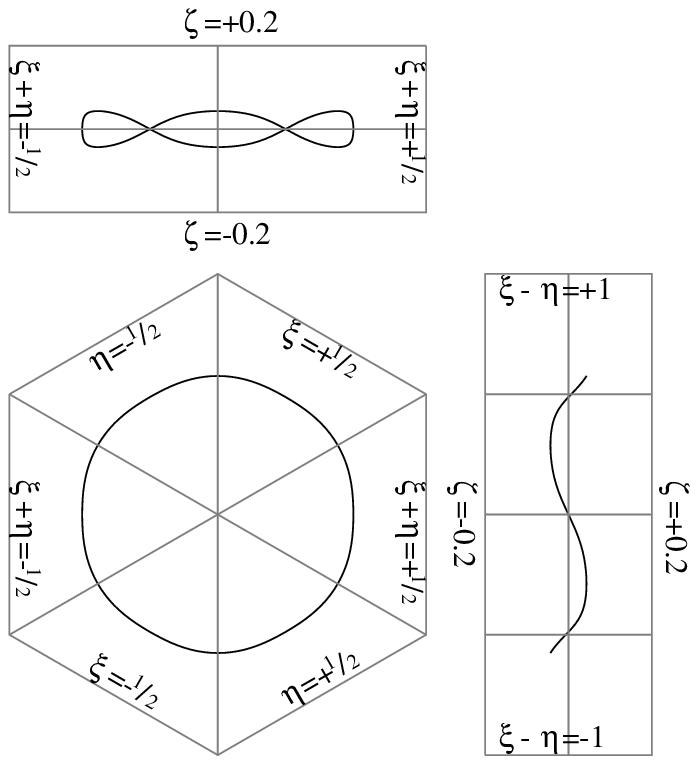}}
\caption{The RUMs given by eq. (\ref{aqu2x}) shown in the $\xi,\eta$-plane and
$\zeta$ versus $\xi+\eta$ and $\xi-\eta$, resp., for
$\alpha$-quartz.\label{aqu2b}}
\end{figure}

(iii) The condition
\be
k_2f_1(\rho_1,\rho_2)+k_3(\rho_3+1/\rho_3)+k_4=0, \quad \cR_2=0 \label{aqu2x}
\ee
yields the curve depicted in fig. \ref{aqu2b}. This curve is not a planar curve.
The projection onto the $\xi,\eta$) plane is nearly a circle. In $\zeta$ direction it oscillates between $-0.0433$ and $+0.0433$. Thus it stays close to the $\zeta=0$ plane. Approximately it can be described as function of the parameter $\chi$
\bea
\xi=r(\chi) \sin(\frac{\pi}6-\chi), & \quad &
\eta=r(\chi) \sin(\frac{\pi}6+\chi), \label{aqu3} \\
r(\chi)=0.3290-0.0035\cos(6\chi), & \quad &
\zeta=0.0452\sin(3\chi)+0.0019\sin(9\chi). \nonumber
\eea
This curve intersects with the lines of RUMs (\ref{aqst0}) at
$\chi=k\pi/3$, and with the RUMs given by the curves (\ref{aqu11}) at $\chi=\pi/6+k\pi/3$ for integer $k$.

\paragraph{Comparison} Comparing the distribution of the RUM-vectors for
$\alpha$-quartz given in \cite{Dove,Dove07} with the curves calculated
here, we find agreement for the straight lines and the curves of figure
(\ref{aqu1}). We do not find the non-planar curve
figure \ref{aqu2b}, eq. (\ref{aqu3}). Since fig. 10 of \cite{Dove07} shows only RUMs for positive $\zeta$, this line should run through the dark regions of quasi-RUMs in the figure.

\end{document}